%% file: main.tex
\documentclass[journal,12pt,draftclsnofoot,onecolumn,twoside]{IEEEtran}

\usepackage{ifpdf}
 \ifpdf
     \usepackage[pdftex]{graphicx}
 \else
     \usepackage[dvips]{graphicx}
 \fi

\graphicspath{{figures/}}

\usepackage{cite}
\usepackage[cmex10]{amsmath}
\usepackage{amssymb,amsfonts}
\usepackage{algorithm}
\usepackage{algpseudocode}

\usepackage{mdwmath}
\usepackage{mdwtab}

\usepackage{flushend}
\usepackage[belowskip=-12pt,aboveskip=-2pt,small]{caption}
\usepackage{setspace}
\begin{document}
%
\title{Optimal Bit and Power Loading for OFDM Systems with Average BER and Total Power Constraints}

%
%
%
\author{{Ebrahim Bedeer, 
Octavia A. Dobre, 
Mohamed H. Ahmed, and 
Kareem E. Baddour \IEEEauthorrefmark{2}}\\
\IEEEauthorblockA{Faculty of Engineering and Applied Science, Memorial University of Newfoundland,
St. John's, NL, Canada\\ 
\IEEEauthorrefmark{2} Communications Research Centre, Ottawa, ON, Canada\\
Email: \{e.bedeer, odobre, mhahmed\}@mun.ca, kareem.baddour@crc.ca}
}
\maketitle

\begin{abstract}
In this paper, a novel joint bit and power loading algorithm is proposed for orthogonal frequency division multiplexing (OFDM) systems operating in fading environments. The algorithm jointly maximizes the throughput and minimizes the transmitted power, while guaranteeing a target average bit error rate (BER) and meeting a constraint on the total transmit power. Simulation results are described that illustrate the performance of the proposed scheme and demonstrate its superiority when compared to the algorithm in \cite{wyglinski2005bit}.
\end{abstract}


%

\input{intro}

\input{opt}
\input{sim}

\input{conclusion}

\section*{Acknowledgment}

This work has been supported in part by the Communications Research Centre, Canada.%






\end{document}

%% file: intro.tex
\section{Introduction}
\vspace{-2.5pt}
Orthogonal frequency division multiplexing (OFDM) modulation represents a robust and efficient transmission technique being adopted by several wireless communication standards \cite{fazel2008multi, mahmoud2009ofdm}. The OFDM system performance can be significantly improved by dynamically adapting the transmission parameters, such as power, constellation size, symbol rate, and coding rate/scheme, according to the channel conditions and the wireless standard specifications \cite{chow1995practical, liu2009adaptive, leke1997maximum, wyglinski2005bit, song2002joint, chung2001degrees, bedeer2012adaptiveRWS,  bedeer2012novelICC}.

Bit and power loading algorithms can be generally categorized into two main classes, i.e., algorithms whose objective is to maximize the achievable system margin, \textit{margin maximization} (MM), \cite{chow1995practical, liu2009adaptive} and algorithms whose objective is to maximize the achievable data rate, \textit{rate maximization} (RM), \cite{leke1997maximum, wyglinski2005bit}.
Most of the techniques proposed in the literature focused on maximizing either the RM or the MM problem separately.
In \cite{chow1995practical}, Chow \textit{et al.} proposed a practical iterative bit loading algorithm to maximize the margin. The algorithm computes the initial bit allocation based on a channel capacity approximation assuming uniform power loading. Then, it iteratively changes the allocated bits to achieve the optimal margin and the target data rate.
Liu and Tang \cite{liu2009adaptive} proposed a low complexity power loading algorithm with uniform bit loading that aims to minimize the transmit power while guaranteeing a target BER. On the other hand, Leke and Cioffi \cite{leke1997maximum} proposed a finite granularity algorithm that maximizes the data rate for a given margin. Subcarriers with signal-to-noise ratio (SNR) below a predefinied threshold are nulled, then remaining subcarriers are identified and the available power is distributed either optimally using a water-filling approach or suboptimally by assuming equal power to maximize the data rate. In \cite{wyglinski2005bit}, Wyglinski \textit{et al.}  proposed an incremental optimal bit loading algorithm with uniform power in order to maximize the throughput, while guaranteeing a target BER. 
Song \textit{et~al.} \cite{song2002joint} proposed an iterative joint bit and power loading algorithm based on statistical channel conditions to meet a target BER.
This algorithm attains a marginal performance improvement when compared to the conventional OFDM systems. The authors conclude that their algorithm is not meant to compete with its counterparts that adapt according the instantaneous channel conditions.
In \cite{bedeer2012novelICC}, the authors proposed a non-iterative low complexity optimal allocation algorithm that jointly maximizes the throughput and minimizes the transmit power, while guaranteeing a target BER per subcarrier.

Emerging wireless communication systems operate under diverse conditions, with different requirements. For example, when operating in interference-prone  shared spectrum environments or in proximity to other frequency-adjacent users, power minimization is crucial. On the other hand, if sufficient guard bands exist to separate users, more emphasis can be given to maximizing the throughput. This motivates us to jointly consider the rate and margin maximization problems. The importance of the competing throughput and power objectives is reflected through a weighting factor.



A novel optimal bit and power loading algorithm is proposed in this paper, which maximizes the throughput and minimizes the total transmit power, subject to  average BER and transmit power constraints. Limiting the total transmit power reduces the interference to existing users, which is crucial in various wireless networks, including cognitive radio environments. Moreover, by including the sum of subcarrier powers in the objective function, the transmit power is minimized even when the power constraint is ineffective, which occurs at smaller signal-to-noise ratios (SNR). Simulation results show that the proposed algorithm outperforms Wyglinski's algorithm \cite{wyglinski2005bit}.

The remainder of the paper is organized as follows. Section \ref{sec:opt} introduces the proposed optimal loading algorithm. Simulation results are presented in Section \ref{sec:sim}, while conclusions are drawn in Section \ref{sec:conc}.

Throughout this paper we use bold-faced upper case letters to denote matrices, e.g., $\mathbf{X}$, bold-faced lower case letters for vectors, e.g., $\mathbf{x}$, and light-faced letters for scalar quantities, e.g., $x$. $\mathbf{I}$ represents the identity matrix, $[.]^T$ denotes the transpose operation, $\nabla$ represents the gradient, and $\lfloor x \rfloor$ is the largest integer not greater than $x$.

%% file: opt.tex
\section{Proposed Algorithm} \label{sec:opt}
\subsection{Optimization Problem Formulation}
An OFDM system decomposes the signal bandwidth into a set of $N$ orthogonal narrowband subcarriers of equal bandwidth. Each subcarrier $i$ transmits $b_i$ bits using power $\mathcal{P}_i$, $i = 1, ..., N$. A delay- and error-free feedback channel is assumed to exist between the transmitter and receiver for reporting channel state information.

In order to minimize the total transmit power and maximize the throughput subject to an average BER and a total power constraint, the optimization problem is  formulated as
\setlength{\arraycolsep}{0.0em}
\begin{equation}
\underset{\mathcal{P}_i}{\textup{Minimize}} \quad \mathcal{P}_T=\sum_{i = 1}^{N}\mathcal{P}_i \quad \textup{and} \quad \underset{b_i}{\textup{Maximize}} \quad b_T=\sum_{i = 1}^{N}b_i, \nonumber
\end{equation} \vspace{-10pt}
\begin{eqnarray}
\textup{subject to} \quad \textup{BER}_{av} = \frac{\sum_{i=1}^{N}b_i \: \textup{BER}_i}{\sum_{i=1}^{N}b_i} &{} \leq {}& \textup{BER}_{th}, \nonumber \\
\sum_{i = 1}^{N}\mathcal{P}_i &{} \leq {}& \mathcal{P}_{th}, \label{eq:eq_first}
\end{eqnarray}
where $\mathcal{P}_T$ and $b_T$ are the total transmit power and throughput, respectively, $\mathcal{P}_{th}$ is the threshold value of the total transmit power, and $\textup{BER}_{av}$, $\textup{BER}_{th}$, and $\textup{BER}_i$ are the average BER, threshold value of BER, and the BER per subcarrier $i$, $i$ = 1, ..., $N$, respectively. An approximate expression for the BER per subcarrier $i$ in the case of $M$-ary QAM is given by\footnote[1]{This expression is tight within 1 dB for BER $\leq$ $10^{-3}$ \cite{chung2001degrees}.} \cite{liu2009adaptive,chung2001degrees}
\setlength{\arraycolsep}{0.0em}
\begin{eqnarray}
\textup{BER}_i &{} \approx  {}& 0.2 \: \textup{exp}\left ( - 1.6 \: \frac{\mathcal{P}_i}{2^{b_i} - 1} \: \frac{\left | \mathcal{H}_i \right |^2}{\sigma^2_n} \right ), \label{eq:BER}
\end{eqnarray}
where $ \mathcal{H}_i $ is the channel gain of subcarrier $i$ and $\sigma^2_n$ is the variance of the additive white Gaussian noise (AWGN).

The multi-objective optimization function in (\ref{eq:eq_first}) can be rewritten as a linear combination of multiple objective functions as follows
\setlength{\arraycolsep}{0.0em}
\begin{equation}
\underset{\mathcal{P}_i , b_i}{\textup{Minimize}}  \quad  \mathcal{F}(\mathbf{p},\mathbf{b}) = \: \left\{\alpha \sum_{i = 1}^{N}\mathcal{P}_i - (1-\alpha)\sum_{i = 1}^{N}b_i\right\}, \label{eq:p1} \nonumber
\end{equation} \vspace{-10pt}
\begin{eqnarray}
\textup{subject to} \quad g_j(\mathbf{p},\mathbf{b}) \leq 0, \quad j = 1,2, \label{eq:ineq_const}
\end{eqnarray}
where $\alpha$ ($0 < \alpha < 1$) is a constant whose value indicates the relative importance of one objective function relative to the other, $\mathbf{p} = [\mathcal{P}_1,...,\mathcal{P}_N]^T$ and $\mathbf{b} = [b_1,...,b_N]^T$ are the \textit{N}-dimensional power and bit distribution vectors, respectively, and $g_j(\mathbf{p},\mathbf{b})$ is the set of constraints given by
\setlength{\arraycolsep}{0.0em}
\begin{equation}
g_j(\mathbf{p},\mathbf{b}) = \left\{\begin{matrix}
0.2 \: \sum_{i=1}^{N} b_i \: \textup{exp}\left ( \frac{- 1.6 \: \mathcal{C}_i \mathcal{P}_i}{2^{b_i} - 1} \right ) - \textup{BER}_{th} \sum_{i=1}^{N} b_i \\ \hspace{5cm} \leq 0, \hfill j = 1  \\
 \sum_{i = 1}^{N}\mathcal{P}_i - \mathcal{P}_{th} \leq 0, \hfill j = 2
\end{matrix}\right.
\label{eq:cons}
\end{equation}
where $\mathcal{C}_i = \frac{\left | \mathcal{H}_i \right |^2}{\sigma^2_n}$ is the channel-to-noise ratio for subcarrier~$i$.

\subsection{Optimization Problem Analysis and Solution}
The problem in (\ref{eq:ineq_const}) can be solved by applying the method of Lagrange multipliers. Accordingly,  the inequality constraints in (\ref{eq:cons}) are transformed to equality constraints by adding non-negative slack variable, $\mathcal{Y}^2_j$, $j$ = 1, 2 \cite{griva2009linear, rao2009engineering}. Hence, the constraints are rewritten as
\setlength{\arraycolsep}{0.0em}
\begin{eqnarray}
\mathcal{G}_j(\mathbf{p},\mathbf{b},\textbf{y}) &{} = {}& g_j(\mathbf{p},\mathbf{b}) + \mathcal{Y}^2_j = 0, \quad j = 1, 2,
\label{eq:eq_const}
\end{eqnarray}
where $\textbf{y} = [\mathcal{Y}_j^2]^T$, $j$ = 1, 2, is the vector of slack variables.  The Lagrange function $\mathcal{L}$ is then expressed as
\setlength{\arraycolsep}{0.0em}
\begin{eqnarray}
\mathcal{\mathcal{L}}(\mathbf{p},\mathbf{b}, \textbf{y}, \boldsymbol\lambda) &{} = {}& \alpha \sum_{i = 1}^{N}\mathcal{P}_i - (1-\alpha)\sum_{i = 1}^{N}b_i \nonumber \\
& + &  \lambda_1\:\Bigg[ 0.2 \: \sum_{i=1}^{N}b_i \: \textup{exp}\left ( \frac{- 1.6 \: \mathcal{C}_i \mathcal{P}_i}{2^{b_i} - 1} \right ) \nonumber \\
& & \hspace{1.8cm} - \: \textup{BER}_{th} \: \sum_{i=1}^{N}b_i + \mathcal{Y}^2_1 \Bigg] \nonumber \\
& + &  \lambda_2\: \Bigg[ \sum_{i = 1}^{N}\mathcal{P}_i - \mathcal{P}_{th} + \mathcal{Y}_2^2  \Bigg],
\end{eqnarray}
where $\boldsymbol\lambda = [\lambda_j]^T$, $j$ = 1, 2, is the vector of Lagrange multipliers. A stationary point can be found when $\nabla \mathcal{L}(\mathbf{p},\mathbf{b},\textbf{y}, \boldsymbol\lambda) = 0$, which yields
\begin{eqnarray}
\frac{\partial \mathcal{L}}{\partial \mathcal{P}_i} &{} = {}& \alpha - 0.2 \times 1.6 \: \lambda_1 \frac{b_i \: \mathcal{C}_i}{2^{b_i}-1} \: \textup{exp}\left ( \frac{- 1.6 \: \mathcal{C}_i \mathcal{P}_i}{2^{b_i} - 1} \right ) \nonumber \\ & & \hspace{5cm} + \lambda_2 = 0,\label{eq:eq1}\\
\frac{\partial \mathcal{L}}{\partial b_i} &{} = {}& -(1 - \alpha) + \lambda_1 \Bigg[0.2 \: \textup{exp}\left ( \frac{- 1.6 \: \mathcal{C}_i \mathcal{P}_i}{2^{b_i} - 1} \right ) \nonumber \\ & &  \bigg[1+ 1.6 \times \ln(2) \frac{\mathcal{C}_i \mathcal{P}_i b_i 2^{b_i}}{(2^{b_i}-1)^2}\bigg] - \textup{BER}_{th} \Bigg] = 0, \label{eq:eq2}\\
\frac{\partial \mathcal{L}}{\partial \lambda_1} & = & 0.2 \: \sum_{i=1}^{N}b_i \: \textup{exp}\left ( \frac{- 1.6 \: \mathcal{C}_i \mathcal{P}_i}{2^{b_i} - 1} \right ) - \: \textup{BER}_{th} \: \sum_{i=1}^{N}b_i \nonumber \\ & & \hspace{5cm}+ \mathcal{Y}^2_1 = 0, \label{eq:eq3}\\
\frac{\partial \mathcal{L}}{\partial \lambda_2} & = & \sum_{i = 1}^{N}\mathcal{P}_i - \mathcal{P}_{th} + \mathcal{Y}_2^2 = 0, \label{eq:eq3_1}\\
\frac{\partial \mathcal{L}}{\partial \mathcal{Y}_1} & = & 2\lambda_1 \mathcal{Y}_1 = 0, \label{eq:eq4} \\
\frac{\partial \mathcal{L}}{\partial \mathcal{Y}_2} & = & 2\lambda_2 \mathcal{Y}_2 = 0. \label{eq:eq4_1}
\end{eqnarray}
It can be seen that (\ref{eq:eq1}) to (\ref{eq:eq4_1}) represent $2N+4$ equations in the $2N+4$ unknowns $\mathbf{p}, \mathbf{b}$, $\textbf{y}$, and $\boldsymbol\lambda$. Equation (\ref{eq:eq4}) implies that either $\lambda_1 = 0$ or $\mathcal{Y}_1 = 0$, while (\ref{eq:eq4_1}) implies that either $\lambda_2$ = 0 or $\mathcal{Y}_2$ = 0. Accordingly, four possible solutions exist, as follows:

--- \textit{Solutions \MakeUppercase{\romannumeral 1} \& \MakeUppercase{\romannumeral 2}}: Choosing $\lambda_1 = 0$ and $\lambda_2$ or $\mathcal{Y}_2$ = 0, results in an underdetermined system of 2 equations in 2$N$+2 unknowns; hence no unique solution can be reached.


--- \textit{Solutions \MakeUppercase{\romannumeral 3} \& \MakeUppercase{\romannumeral 4}}: Choosing $\mathcal{Y}_1$ = 0 and $\lambda_2$ = 0 (the total power constraint is inactive) or $\mathcal{Y}_2$ = 0 (the total power constraint is active), we obtain a system $\mathcal{S(\mathbf{x})}$ of $2N+2$ equations in the  $2N+2$ unknowns $\mathbf{x}$, where $\mathbf{x} = [\mathbf{p}, \mathbf{b}, \lambda_1, \mathcal{Y}_2]$, that cannot be solved analytically. Hence, we resort to solve this system numerically. Various numerical methods are available in the literature, e.g., the steepest descent, the Gauss-Newton, and the Levenberg-Marquardt (LM) methods  \cite{rao2009engineering, madsen1999methods}. The steepest descent method is efficient when $\mathbf{x}$ is far from the optimal solution $\mathbf{x}_{op}$. On the other hand, the Gauss-Newton method converges fast when $\mathbf{x}$ is close to $\mathbf{x}_{op}$. The LM method takes advantage of both methods by introducing a positive damping factor $\mu_k$ to control the step size at every iteration $k$ depending on the closeness to $\mathbf{x}_{op}$.


The LM algorithm is briefly discussed here for completeness of the presentation; however, further details can be found in \cite{rao2009engineering, madsen1999methods}. We start from an initial point $\mathbf{x}_0$ and initial step $\mathbf{d}_0$, then a series of points $\mathbf{x}_1$, $\mathbf{x}_2$, .... is obtained that converges towards the solution $\mathbf{x}_{op}$; hence, at iteration $k$ one can write  $\mathbf{x}_{k+1} = \mathbf{x}_k +  \mathbf{d}_k$, where $\mathbf{d}_k$ is
the LM step given by \cite{rao2009engineering, madsen1999methods}
\begin{eqnarray}
\mathbf{d}_k & = & - \big[\textbf{J}(\mathbf{x}_k)^T \: \textbf{J}(\mathbf{x}_k) + \mu_k \mathbf{I} \big]^{-1} \textbf{J}(\mathbf{x}_k)^T\mathcal{S}(\mathbf{x}_k), \label{eq:LM_step}
\end{eqnarray}
where $\mathbf{I}$ is the identity matrix and $\textbf{J}(\mathbf{x}_k)$ is the Jacobian matrix of the system $\mathcal{S}(\mathbf{x}_k)$, defined earlier, both at point $\mathbf{x}_k$. The damping parameter $\mu_k$ has several advantages. First, for all $\mu_k > 0$, the matrix $\textbf{J}(\mathbf{x}_k)^T \: \textbf{J}(\mathbf{x}_k) + \mu_k \mathbf{I}$ is positive definite, which insures that $\mathbf{d}_k $ has a descent direction and that the system $\mathcal{S(\mathbf{x})}$ has a unique solution. Second, if $\mu_k$ is large, the step value is given by $\mathbf{d}_k  \simeq - \frac{1}{\mu_k} \textbf{J}(\mathbf{x}_k)^T\mathcal{S}(\mathbf{x}_k)$ representing a short step in the steepest descent direction, and is preferred if the current iteration is far from $\mathbf{x}_{op}$. On the other hand, if $\mu_k$ is very small, then $\mathbf{d}_k$ equals the Gauss-Newton step which is suitable in the final stages of the iterations, i.e., when $\mathbf{x}_k$ is close to $\mathbf{x}_{op}$. Third, it prevents the step $\mathbf{d}_k $ from being too large when $\textbf{J}(\mathbf{x}_k)^T \: \textbf{J}(\mathbf{x}_k)$ is nearly singular. Furthermore, it guarantees that the step is defined when $\textbf{J}(\mathbf{x}_k)^T \: \textbf{J}(\mathbf{x}_k)$ is singular, in contrast to the Gauss-Newton method where the step is undefined.


%

\subsection{Description of the Proposed Algorithm}
To solve the problem defined in (\ref{eq:ineq_const}), we propose the following algorithm. Given an initial point $\mathbf{x}_0$, the value of $\mathcal{S}(\mathbf{x}_0)$ is calculated, and the initial step $\mathbf{d}_0$ is determined according to (\ref{eq:LM_step}), then we set $\mathbf{x}_1 = \mathbf{x}_0 + \mathbf{d}_0$, and the process repeats.  At each iteration $k$, if $\mu_k$ is large, i.e., small $\mathbf{d}_k $ step, then $\mu_{k+1}$ is decreased to approximate the Gauss-Newton step and converges faster to $\mathbf{x}_{op}$; otherwise $\mu_{k+1}$ is increased to approximate a steepest descent step. The algorithm converges to the optimal solution $\mathbf{x}_{op}$ at iteration $k$ if \textit{both} $\mathcal{S}(\mathbf{x}_k)$ and $\mathbf{d}_k $ are less than the tolerance errors  $\epsilon$ and $\varepsilon$, respectively\footnote[2]{If either $\mathcal{S}(\mathbf{x}_0) < \epsilon$ \textit{or} $\mathbf{d}_0 < \varepsilon$, the algorithm stops without converging.}. To avoid an infinite loop, we set the maximum allowed number of iterations to $k_{max}$ (if the number of iterations reach $k_{max}$, this means that the algorithm could not converge to the optimal solution $\mathbf{x}_{op}$). Once $\mathbf{x}_{op}$ is reached, $\mathbf{p}_{op}$ and $\mathbf{b}_{op}$ are obtained and the final bit and power distributions are calculated by rounding down the non-integer $\mathbf{b}_{op}$, while keeping the power distribution the same, i.e., $\mathbf{b}_{final} = \lfloor \mathbf{b}_{op} \rfloor$ and $\mathbf{P}_{final} = \mathbf{P}_{op}$. The proposed algorithm can be formally stated as follows.
\floatname{algorithm}{}
\begin{algorithm}
\renewcommand{\thealgorithm}{}
\caption{\textbf{Proposed Algorithm}}
\begin{algorithmic}[1]
\State \textbf{INPUT} The AWGN variance $\sigma^2_n$, channel gain per subcarrier $i$ ($\mathcal{H}_i$), threshold value of average BER ($\textup{BER}_{th}$), threshold value of the total transmit power $\mathcal{P}_{th}$, weighting factor $\alpha$, $\nu_1$ ($0 < \nu_1 < 1$), $\nu_2$ ($\nu_2 > 1$), and tolerance errors $\varepsilon$, and $\epsilon$.
\State Set the iteration number $k$ to 0.
\State Pick an initial solution $\mathbf{x}_0$ and initial damping parameter $\mu_0$.
%
\While {$\mathcal{S}(\mathbf{x}_k) > \epsilon$ and $\mathbf{d}_k  > \varepsilon$ and $k < k_{max}$}
\State $k = k +1$
\State $\mathbf{d}_k  = - \big[\textbf{J}(\mathbf{x}_k)^T \: \textbf{J}(\mathbf{x}_k) + \mu_k \mathbf{I} \big]^{-1} \textbf{J}(\mathbf{x}_k)^T\mathcal{S}(\mathbf{x}_k)$
\State $\mathbf{x}_{op} = \mathbf{x}_k +  \mathbf{d}_k $
\If {$\mu_k > \mu_{th}$\footnotemark[3]}
\State $\mathbf{x}_{k+1} = \mathbf{x}_{op}$
\State $\mu_{k+1} = \nu_1 \: \mu_k$
\Else
\State $\mu_{k+1} = \nu_2 \: \mu_k$
\EndIf
\EndWhile
\State Given $\mathbf{x} = [\mathbf{p}, \mathbf{b}, \lambda]$, find the values of $\mathbf{p}_{op}$ and $\mathbf{b}_{op}$ corresponding to $\mathbf{x}_{op}$.
\For{$i$ = 1, ..., $N$}
\If{$b_{i,op} \geq$ 2}
\State $b_{i,final} = \lfloor b_{i,op} \rfloor$ and $\mathcal{P}_{i,final} = \mathcal{P}_{i,op}$
\Else
\State $b_{i,final}$ and $\mathcal{P}_{i,final}$ = 0
\EndIf
\EndFor
\State \textbf{OUTPUT} $b_{i,final}$ and $\mathcal{P}_{i,final}$, $i$ = 1, ..., $N$.
\end{algorithmic}
\end{algorithm}
\footnotetext[3]{For more details on the choice of $\mu_{th}$ we refer the reader to \cite{rao2009engineering, madsen1999methods}.}

%% file: sim.tex
\section{Numerical Results} \label{sec:sim}

This section investigates the performance of the proposed algorithm in terms of the achieved average throughput and average transmit power, and compares its performance with the algorithm in \cite{wyglinski2005bit}. 

\subsection{Simulation Setup}
An OFDM system with a total of $N$ = 128 subcarriers is considered. The channel impulse response $h(n)$ of length $N_{ch}$ is modeled as independent complex Gaussian random variables with zero mean and exponential power delay profile \cite{lin2008low}   
\setlength{\arraycolsep}{0.0em}
\begin{eqnarray}
\mathbb{E}\{\left | h(n) \right |^2\} = \sigma_h^2 \: e^{-n\Xi}, \qquad n = 0, 1, ..., N_{ch}-1,
\end{eqnarray}
where $\sigma_h^2$ is a constant chosen such that the average energy per subcarrier is normalized to unity, i.e., $\mathbb{E}\{\left | \mathcal{H}_i \right |^2\}$ = 1, and $\Xi$ represents the decay factor. Representative results are presented in this section and were obtained by repeating Monte Carlo trials for $10^{3}$ channel realizations with a channel length $N_{ch} = 5$ taps, decay factor  $\Xi = \frac{1}{5}$, $\textup{BER}_{th} = 10^{-4}$. The LM algorithm parameters are as follows: $\mu_0 = 10^5$, $\nu_1 = 0.5$, $\nu_2 = 2$, $\varepsilon = \epsilon = 10^{-6}$,  and $k_{max} = 10^4$. 

\subsection{Performance of the Proposed Algorithm}

\begin{figure}[!t]
	\centering
		\includegraphics[width=0.50\textwidth]{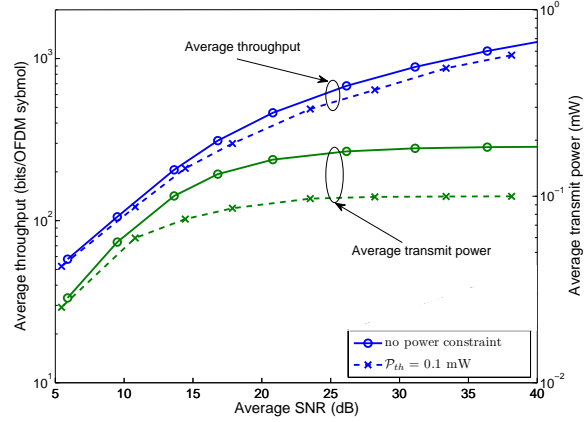}
	\caption{Average throughput and average transmit power as a function of average SNR, with and without power constraint, at $\alpha$ = 0.5.}
	\label{fig:proposed}
\end{figure}

Fig. \ref{fig:proposed} depicts the average throughput and average transmit power as a function of the average SNR\footnote[4]{The average SNR is calculated by averaging the instantaneous SNR values per subcarrier over the total number of subcarriers and the total number of channel realizations, respectively.}  at $\alpha$  = 0.5, with and without considering the total power constraint. Without considering the total power constraint and for an average SNR $\leq$ 21 dB, one finds that both the average throughput and the average transmit power increase as the SNR increases, whereas for an average SNR $\geq$ 21 dB, the transmit power saturates, and the throughput continues to increase. This observation can be explained as follows. For lower values of the average SNR, many subcarriers are nulled. By increasing the average SNR, the number of used subcarriers increases, resulting in a noticeable increase in the throughput and power. Apparently, for average SNR $\geq$ 21 dB, all subcarriers are used, and our proposed algorithm essentially minimizes the average transmit power by keeping it constant, while increasing the average throughput. When a power constraint $\mathcal{P}_{th}$ = 0.1 $mW$ is considered, at lower SNR values the total transmit power is below this threshold and both the allocated power and throughput levels are similar to the no constraint case. However, at higher SNR values, when the total transmit power exceeds the threshold, a small reduction in the average throughput is noticed, which emphasizes that the proposed algorithm meets the power constraint while maximizing the throughput, i.e., the throughput does not degrade much when compared to the case of no power constraint.

\begin{figure}[!t]
	\centering
		\includegraphics[width=0.50\textwidth]{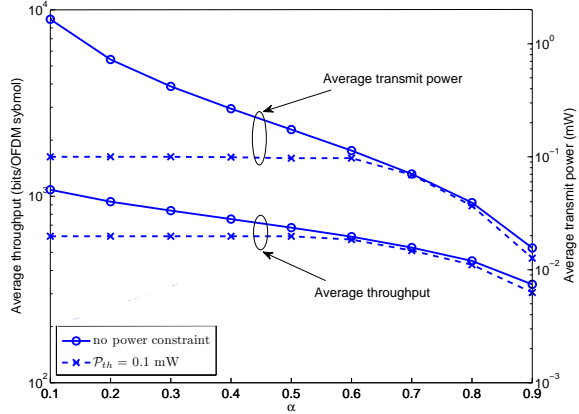}
	\caption{Average throughput and average transmit power as a function of weighting factor $\alpha$, with and without power constraint, at $\sigma_n^2 = 10^{-3}$ $\mu W$.}
	\label{fig:proposed_alpha}
\end{figure}

In Fig. \ref{fig:proposed_alpha}, the average throughput and average transmit power are plotted as a function of the weighting factor $\alpha$  at  $\sigma_n^2 = 10^{-3}$ $\mu$W, with and without considering the total power constraint. Without considering the total power constraint, one can notice that an increase of the weighting factor $\alpha$ yields a decrease of both the average throughput and average transmit power. This can be explained as follows. By increasing $\alpha$, more weight is given to the transmit power  minimization (the minimum transmit power is further reduced), whereas less weight is given to the throughput maximization (the maximum throughput is reduced), according to the problem formulation. By considering a total power constraint, $\mathcal{P}_{th}$ = 0.1 $mW$, the same average throughput and power are obtained if the total transmit power is less than $\mathcal{P}_{th}$, while the average throughput and power saturate if the total transmit power exceeds $\mathcal{P}_{th}$. Note that this is different from Fig.~\ref{fig:proposed}, where the average throughput increases while the transmit power is kept constant, which is due the increase of the average SNR value. Fig. \ref{fig:proposed_alpha} illustrates the benefit of introducing such a weighting factor in our problem formulation to tune the average throughput and transmit power levels as needed by the wireless communication system.

\begin{figure}[!t]
	\centering
		\includegraphics[width=0.50\textwidth]{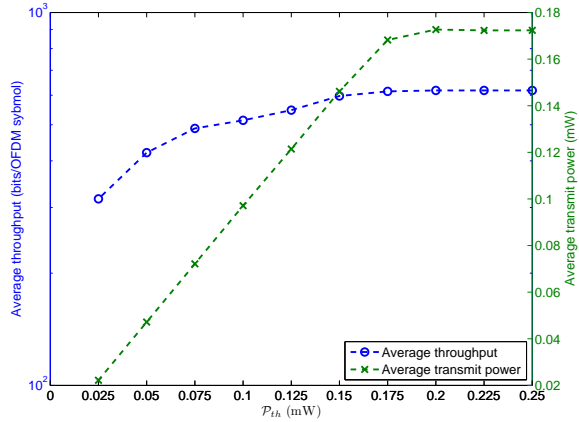}
	\caption{Average throughput and average transmit power as a function of the power constraint $\mathcal{P}_{th}$, at $\alpha$ = 0.5 and $\sigma_n^2 = 10^{-3}$ $\mu W$.}
	\label{fig:T_P_vs_power_constraint}
\end{figure}

In Fig. \ref{fig:T_P_vs_power_constraint}, the average throughput and average transmit power are plotted as a function of the power threshold $\mathcal{P}_{th}$, at $\alpha$ = 0.5 and $\sigma_n^2 = 10^{-3}$ $\mu W$. It can be noticed that the average throughput increases as $\mathcal{P}_{th}$ increases, and saturates for higher values of $\mathcal{P}_{th}$; moreover, the average transmit power increases linearly with $\mathcal{P}_{th}$, while it saturates for higher values of $\mathcal{P}_{th}$. This can be explained, as for lower values of $\mathcal{P}_{th}$, the total transmit power is restricted by this threshold value, while increasing this threshold value results in a corresponding increase in both the average throughput and  total transmit power. For higher values of $\mathcal{P}_{th}$, the total transmit power is always less than the threshold value, and, thus, it is as if the constraint on the total transmit power is actually relaxed. In this case, the proposed algorithm essentially minimizes the transmit power by keeping it constant; consequently, the average throughput remains constant for the same noise variance as in Fig. \ref{fig:proposed_alpha}.

\subsection{Performance Comparison with the Algorithm in \cite{wyglinski2005bit}}

\begin{figure}
	\centering
		\includegraphics[width=0.50\textwidth]{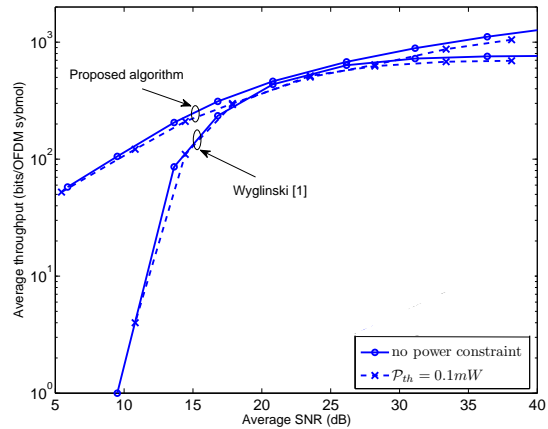}
	\caption{Average throughput as a function of average SNR for the proposed algorithm and Wyglinski's algorithm \cite{wyglinski2005bit}.}
	\label{fig:throughput}
\end{figure}

In Fig. \ref{fig:throughput}, the throughput achieved by the proposed algorithm is compared to that obtained by Wyglinski's algorithm \cite{wyglinski2005bit} for the same operating conditions. To make a fair comparison, the uniform power loading used by the loading scheme in \cite{wyglinski2005bit} is computed by dividing the average transmit power allocated by our algorithm by the total number of subcarriers. As can be seen in Fig.~\ref{fig:throughput}, the proposed algorithm provides a significantly higher throughput than the scheme in \cite{wyglinski2005bit} for low average SNR values. This result demonstrates that optimal loading of transmit power is crucial for low power budgets. 

%% file: conclusion.tex
\section{Conclusion} \label{sec:conc}
In this paper, we proposed a novel algorithm that jointly maximizes the throughput and minimizes the transmit power with constraints on the average BER and the total transmit power, for OFDM systems. Simulation results demonstrated the good performance of the proposed algorithm, which also outperforms the loading algorithm in \cite{wyglinski2005bit} under the same operating conditions.


%% file: main.bbl
\begin{thebibliography}{10}
	\providecommand{\url}[1]{#1}
	\csname url@samestyle\endcsname
	\providecommand{\newblock}{\relax}
	\providecommand{\bibinfo}[2]{#2}
	\providecommand{\BIBentrySTDinterwordspacing}{\spaceskip=0pt\relax}
	\providecommand{\BIBentryALTinterwordstretchfactor}{4}
	\providecommand{\BIBentryALTinterwordspacing}{\spaceskip=\fontdimen2\font plus
		\BIBentryALTinterwordstretchfactor\fontdimen3\font minus
		\fontdimen4\font\relax}
	\providecommand{\BIBforeignlanguage}[2]{{%
			\expandafter\ifx\csname l@#1\endcsname\relax
			\typeout{** WARNING: IEEEtran.bst: No hyphenation pattern has been}%
			\typeout{** loaded for the language `#1'. Using the pattern for}%
			\typeout{** the default language instead.}%
			\else
			\language=\csname l@#1\endcsname
			\fi
			#2}}
	\providecommand{\BIBdecl}{\relax}
	\BIBdecl
	
	\bibitem{wyglinski2005bit}
	A.~Wyglinski, F.~Labeau, and P.~Kabal, ``Bit loading with {BER}-constraint for
	multicarrier systems,'' \emph{{IEEE} Trans. Wireless Commun.}, vol.~4, no.~4,
	pp. 1383--1387, Jul. 2005.
	
	\bibitem{fazel2008multi}
	K.~Fazel and S.~Kaiser, \emph{Multi-carrier and Spread Spectrum Systems: from
		{OFDM} and {MC-CDMA} to {LTE} and {WiMAX}}.\hskip 1em plus 0.5em minus
	0.4em\relax John Wiley \& Sons Inc, 2008.
	
	\bibitem{mahmoud2009ofdm}
	H.~Mahmoud, T.~Yucek, and H.~Arslan, ``{OFDM} for cognitive radio: merits and
	challenges,'' \emph{{IEEE} Wireless Commun.}, vol.~16, no.~2, pp. 6--15, Apr.
	2009.
	
	\bibitem{chow1995practical}
	P.~Chow, J.~Cioffi, and J.~Bingham, ``A practical discrete multitone
	transceiver loading algorithm for data transmission over spectrally shaped
	channels,'' \emph{{IEEE} Trans. Commun.}, vol.~43, no. 234, pp. 773--775,
	Feb. 1995.
	
	\bibitem{liu2009adaptive}
	K.~Liu, B.~Tang, and Y.~Liu, ``Adaptive power loading based on unequal-{BER}
	strategy for {OFDM} systems,'' \emph{{IEEE} Commun. Lett.}, vol.~13, no.~7,
	pp. 474--476, Jul. 2009.
	
	\bibitem{leke1997maximum}
	A.~Leke and J.~Cioffi, ``A maximum rate loading algorithm for discrete
	multitone modulation systems,'' in \emph{Proc. {IEEE} Global
		Telecommunications Conference ({GLOBECOM})}, vol.~3, Nov. 1997, pp.
	1514--1518.
	
	\bibitem{song2002joint}
	Z.~Song, K.~Zhang, and Y.~Guan, ``Joint bit-loading and power-allocation for
	{OFDM} systems based on statistical frequency-domain fading model,'' in
	\emph{Proc. {IEEE} Vehicular Technology Conference ({VTC})-Fall}, Sep. 2002,
	pp. 724--728.
	
	\bibitem{chung2001degrees}
	S.~Chung and A.~Goldsmith, ``Degrees of freedom in adaptive modulation: a
	unified view,'' \emph{{IEEE} Trans. Commun.}, vol.~49, no.~9, pp. 1561--1571,
	Sep. 2001.
	
	\bibitem{bedeer2012adaptiveRWS}
	E.~Bedeer, M.~Marey, O.~A. Dobre, and K.~Baddour, ``Adaptive bit allocation for
	{OFDM} cognitive radio systems with imperfect channel estimation,'' in
	\emph{Proc. {IEEE} Radio and Wireless Symposium}, Jan. 2012, pp. 359-- 362.
	
	\bibitem{bedeer2012novelICC}
	E.~Bedeer, M.~Marey, O.~Dobre, M.~Ahmed, and K.~Baddour, ``A novel algorithm
	for joint bit and power loading for {OFDM} systems with unknown
	interference,'' in \emph{Proc. {IEEE} International Conference on
		Communications ({ICC})}, Jun. 2012, pp. 3605--3610.
	
	\bibitem{griva2009linear}
	I.~Griva, S.~Nash, and A.~Sofer, \emph{Linear and Nonlinear
		Optimization}.\hskip 1em plus 0.5em minus 0.4em\relax Society for Industrial
	Mathematics, 2009.
	
	\bibitem{rao2009engineering}
	S.~Rao, \emph{Engineering Optimization: Theory and Practice}.\hskip 1em plus
	0.5em minus 0.4em\relax Wiley, 2009.
	
	\bibitem{madsen1999methods}
	K.~Madsen, H.~Bruun, and O.~Tingleff, \emph{Methods for Non-linear Least
		Squares Problems}.\hskip 1em plus 0.5em minus 0.4em\relax Informatics and
	Mathematical Modelling, Technical University of Denmark, 2nd edition, 2004.
	
	\bibitem{lin2008low}
	H.~Lin, X.~Wang, and K.~Yamashita, ``A low-complexity carrier frequency offset
	estimator independent of {DC} offset,'' \emph{{IEEE} Commun. Lett.}, vol.~12,
	no.~7, pp. 520--522, Jul. 2008.
	
\end{thebibliography}
